\documentclass[%
reprint,
superscriptaddress,
 amsmath,amssymb,
 aps,
pra,
]{revtex4-1}

\usepackage{graphicx}
\usepackage{dcolumn}
\usepackage{bm}

\usepackage{amsmath,amssymb}
\usepackage[capitalize]{cleveref}
\usepackage[separate-uncertainty=true,multi-part-units=single]{siunitx}
\usepackage{float}
\DeclareSIUnit\mbar{\milli\bar}

\usepackage{xparse}
\usepackage{xcolor}
\definecolor{mygray}{gray}{0.6}

\begin{document}

\title{Spontaneous electron emission from hot silver dimer anions: Breakdown of the Born-Oppenheimer approximation}

\author{E.~K.~Anderson}
\email{emma.anderson@fysik.su.se}
\affiliation{Department of Physics, Stockholm University, AlbaNova, SE-106 91 Stockholm, Sweden}

\author{A.~F.~Schmidt-May}
\affiliation{Institut f\"ur Ionenphysik und Angewandte Physik, Universit\"at Innsbruck, A-6020 Innsbruck, Austria}
\affiliation{Department of Physics, Stockholm University, AlbaNova, SE-106 91 Stockholm, Sweden}

\author{P.~K.~Najeeb}
\affiliation{Department of Physics, Stockholm University, AlbaNova, SE-106 91 Stockholm, Sweden}

\author{G.~Eklund}
\affiliation{Department of Physics, Stockholm University, AlbaNova, SE-106 91 Stockholm, Sweden}

\author{K.~C.~Chartkunchand}
\affiliation{Atomic, Optical, and Molecular Physics Laboratory, RIKEN Cluster for Pioneering Research Wako-shi, Saitama 351-0198, Japan}

\author{S.~Ros\'en}
\affiliation{Department of Physics, Stockholm University, AlbaNova, SE-106 91 Stockholm, Sweden}

\author{\AA.~Larson}
\affiliation{Department of Physics, Stockholm University, AlbaNova, SE-106 91 Stockholm, Sweden}

\author{K.~Hansen}
\affiliation{Center for Joint Quantum Studies and 
Department of Physics, Tianjin University, 92 Weijin Road, Tianjin 300072, China}
\affiliation{Department of Physics, University of Gothenburg, 41296 Gothenburg, Sweden}

\author{H.~Cederquist}
\affiliation{Department of Physics, Stockholm University, AlbaNova, SE-106 91 Stockholm, Sweden}

\author{H.~Zettergren}
\affiliation{Department of Physics, Stockholm University, AlbaNova, SE-106 91 Stockholm, Sweden}

\author{H.~T.~ Schmidt}
\email{henning.schmidt@fysik.su.se}
\affiliation{Department of Physics, Stockholm University, AlbaNova, SE-106 91 Stockholm, Sweden}

\date{\today}

\begin{abstract}


We report the first experimental evidence of spontaneous electron emission from a homonuclear dimer anion through direct measurements of $\rm{Ag}_2^- \rightarrow \rm{Ag}_2 + \rm{e}^-$ decays on milliseconds and seconds time scales. This observation is very surprising as there is no avoided crossing between adiabatic energy curves to mediate such a process. The process is weak but yet dominates the decay signal after 100\,ms when ensembles of internally hot Ag$_2^-$ ions are stored in the cryogenic ion-beam storage ring, DESIREE, for 10 seconds.
The electron emission process is associated with an instantaneous, very large, reduction of the vibrational energy of the dimer system. This represents a dramatic deviation from a Born-Oppenheimer 
description of dimer dynamics.

\end{abstract}
\maketitle





In the Born-Oppenheimer approximation for molecules, the electronic and nuclear degrees of freedom are treated separately~\cite{Born1927}. This is the basis for representing molecular quantum states as products of electronic and vibrational-rotational wavefunctions. Strictly within such a description, no instantaneous transfer of energy between electronic and nuclear degrees of freedom is possible. For {\it complex} systems, rapid transfers of energy between electronic and nuclear degrees of freedom are nevertheless common.
The energy transfer may happen when such systems are photoexcited and evolve from an electronically excited potential energy surface via one or more conical intersections to the ground state
\cite{Nielsen2001}. The reverse process---inverse internal conversion---where a vibrationally hot complex system in the electronic ground state converts vibrational energy to electronic excitations is possible as well and may lead to recurrent fluorescence processes \cite{Kono2015,Ebara2016,Kono2018,Martin2013}. Rapid transfer of energy between the electronic and nuclear degrees of freedom is also expected for most small metal cluster anions \cite{Anderson2018,Hansen2017,Kafle2015}.  These processes are often very fast. Typically, they occur on picosecond timescales at avoided crossings or conical intersections 
with small separations between potential energy curves or surfaces.
Recently, electron emission from excited sulfur hexafluoride anions was ascribed to a crossing of the anion and neutral potential energy surfaces \citep{Menk2014}, yielding a strong coupling to the electron continuum states.

For a range of homonuclear metal dimer anions such as Cu$_2^-$, Ag$_2^-$, and Au$_2^-$, the situation is fundamentally different as there are no avoided crossings between adiabatic potential energy curves to drive the transition.
Yet, electron emission must be accompanied by a dramatic decrease
in vibrational excitation. The kinetic energies of the two nuclei must be reduced by an amount exceeding the electron affinity of the dimer, which in the case of Ag$_2$ is $EA=1.023\pm0.007$\,eV \cite{Ho1990}. 


In this letter we report a study of the spontaneous uni-molecular decays of internally hot Ag$_2^-$ anions stored in the cryogenic ion-beam storage ring, DESIREE \cite{Schmidt2013,Thomas2011}. In the experiment, we separate the contributions from fragmentation, Ag$_2^-$ $\rightarrow$ Ag + Ag$^-$, and electron emission, Ag$_2^-$ $\rightarrow$ Ag$_2$ + e$^-$. We find that the former process dominates the decay signal up to 100\,ms of storage while the latter process, very surprisingly, dominates the signal thereafter.

\begin{figure}
	\centering
	\includegraphics[width=1\columnwidth]{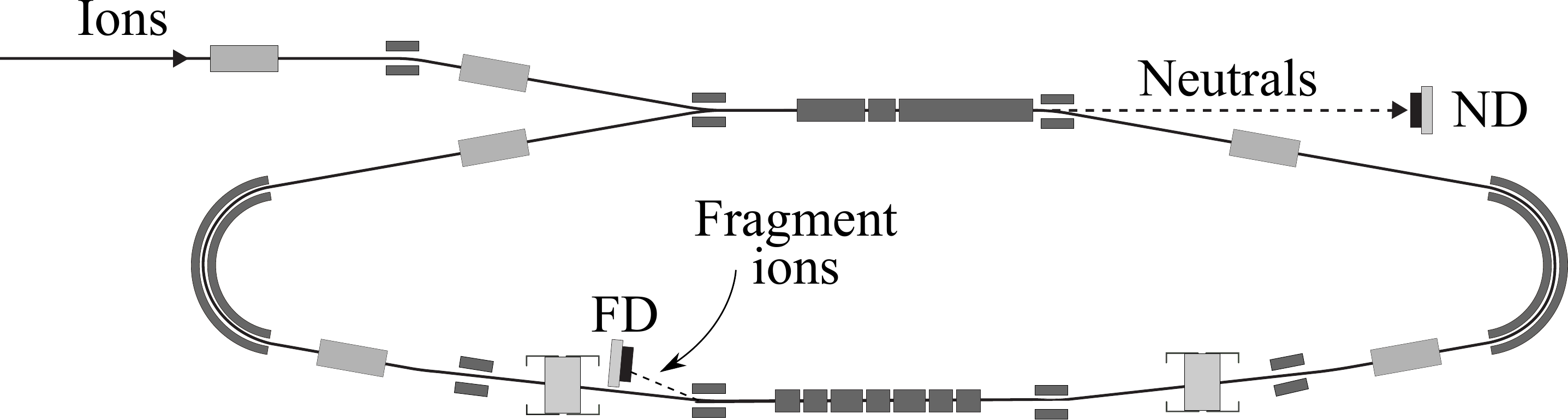}
	\caption{Schematic of the electrostatic ion-beam storage ring used for the present measurements. A 10 keV Ag$_2^-$ silver dimer anion bunch of about 10$^5$ ions is injected to circulate in a closed orbit at a revolution frequency of 10.8\,kHz. Neutral particles Ag and Ag$_2$ from spontaneous fragmentation and from electron detachment processes in the straight section closest to the injection port are detected by the neutral detector (ND). Atomic anions, Ag$^-$, from fragmentation events in the opposite straight section are detected by the fragment detector (FD).} 
	\label{fig:apparatus}
\end{figure}


In Fig.~\ref{fig:apparatus}, we show a schematic of one of the DESIREE \cite{Thomas2011,Schmidt2013} ion-beam storage rings 
with its detectors for neutrals (Neutral Detector, ND) and for fragments (Fragment Detector, FD). The ring is operated at cryogenic temperatures and with a residual-gas density of only a few H$_2$ molecules per mm$^3$.
This enables ion storage for tens of minutes \cite{Backstrom2015}. A cesium sputter ion source with a silver cathode is used to produce silver cluster anions in high vibrational and rotational states. These anions are accelerated to 10\,keV. A 90$^{\circ}$ analyzing magnet is then used to select a 150-200\,pA $^{107}$Ag$_2^-$ beam for injection into the ring, yielding about 10$^5$ stored ions. This low current is used in order to avoid saturation of the detectors.

Fragmentation and electron detachment are the only two pathways for the rotationally and vibrationally excited dimer anions to spontaneously decay.
Ro-vibrationally excited $^{107}$Ag$_2^-$ ions in the electronic ground state cannot relax radiatively through electric-dipole transitions. Higher multipole order decays are possible,  but are expected to be extremely slow---no calculation for Ag$_2^-$ exists but for another homonuclear dimer, H$_2^+$, the corresponding lifetimes have been calculated to be of the order of {\it days} \cite{Peek1979}. 
Consistent with these expectations, no radiative cooling is observed for the silver dimer anions in the present experiment.

The rates of neutral products (Ag$_2$, Ag) from fragmentation and electron detachment events in one of the straight sections of the ring (Fig.\ref{fig:apparatus}) are measured as functions of time after ion production using the ND detector. The FD detector is used to measure the rate of Ag$^-$ products from fragmentation events (Fig.\ref{fig:apparatus}). All data is recorded in event-by-event mode with the time after ion formation in the source recorded for each count on either detector. 

\begin{figure}
	\centering
	\includegraphics[width=1\columnwidth]{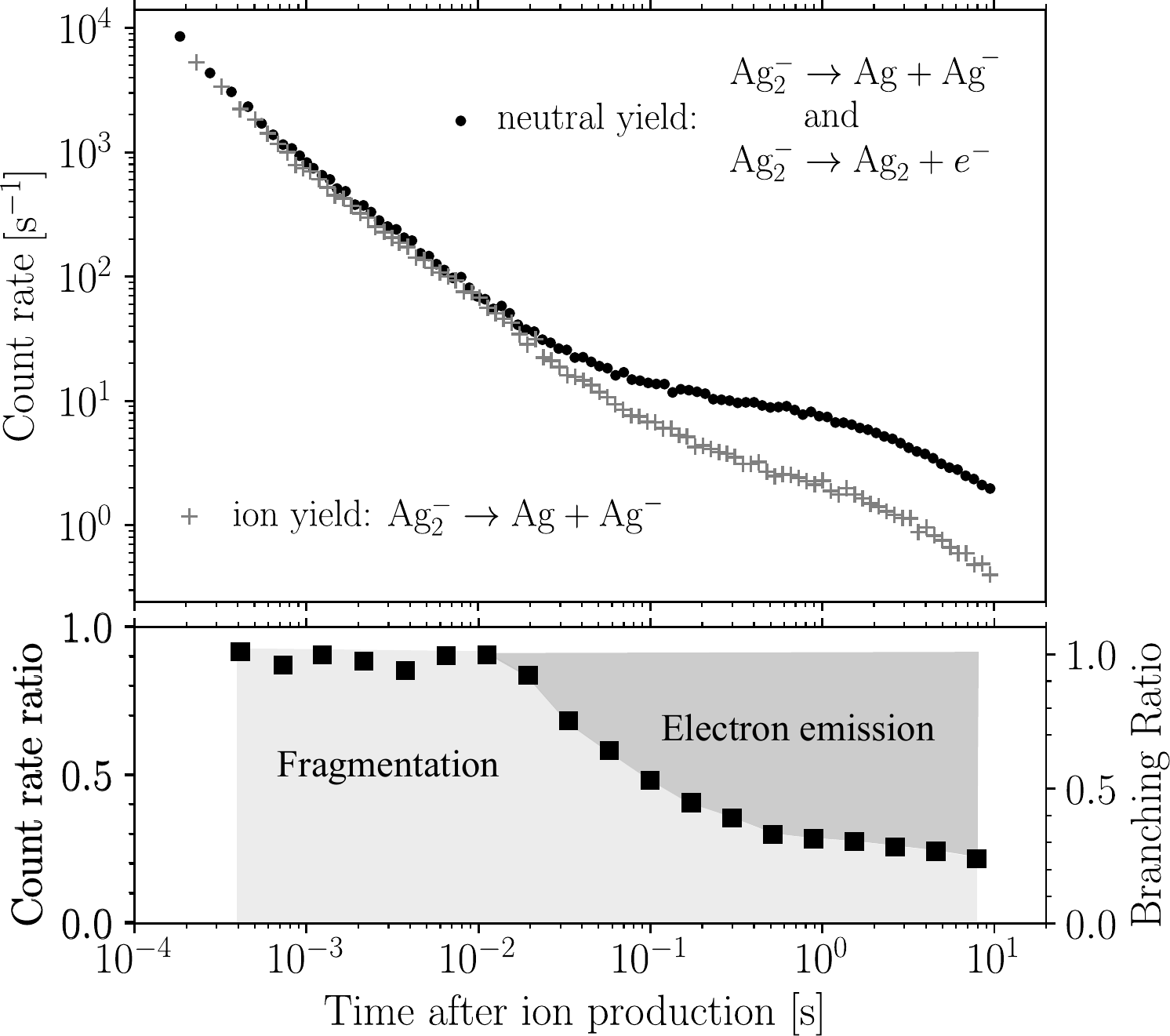}
	\caption{Data recorded with stored 10 keV Ag$_2^-$ anions. The upper panel shows detector count rates as functions of time after the ion production in the Cs sputter ion source: ${\bullet}$ Background corrected count rates of neutral products recorded by the neutral detector (ND in Fig.\,\ref{fig:apparatus}). $\textcolor{mygray}{+}$ Ionic Ag$^-$ fragments recorded by the fragment detector (FD in Fig.\,\ref{fig:apparatus}). Lower panel: The ratio 
between the Ag$^-$ and (Ag + Ag$_2$) count rates as measured by the FD and ND detectors, respectively (left vertical axis). The vertical axis to the right has been corrected for the slightly different detection efficiencies of the two detectors (see text) and thus shows the branching ratio for fragmentation directly. The statistical errors are smaller than the symbols in both panels.} 
	\label{fig:Ag2_Ratios}
\end{figure}

The measured count rates of neutrals (Ag from fragmentation and Ag$_2$ from electron emission) and fragment ions (Ag$^-$ from fragmentation) leaving the ensembles of stored Ag$_2^-$ ions are shown in the upper panel of Fig.~\ref{fig:Ag2_Ratios}. The FD and ND detectors are equivalent; both are 40\,mm diameter microchannel-plate detector assemblies with resistive anodes. For this reason, the two count rates are directly displayed without any normalization in the upper panel of Fig.~\ref{fig:Ag2_Ratios}. The zero point for the time scale is set to be the moment of formation of the ions in the cesium sputter ion source, and the data is binned with bin widths increasing linearly with time for clear presentation on a logarithmic scale with the exception of the first few data points which each represent single revolutions of 93\,$\rm{\mu}$s duration. The detector dark count background rates are measured prior to ion beam injection to be 6.27$\pm$0.03\,s$^{-1}$ and 4.32$\pm$0.03\,s$^{-1}$ for the neutral and fragment detectors, respectively, and have been subtracted from the data shown in Fig.~\ref{fig:Ag2_Ratios}. Up to 10\,ms after the ensemble of stored $^{107}$Ag$_2^-$ ions were formed, we see that the two detectors have very similar count rates and identical time dependences. We take this as evidence that the fragmentation process is completely dominating yielding equal numbers of neutral-atom and atomic-anion products within this time range.

In the lower panel of Fig.~\ref{fig:Ag2_Ratios}, we show the ratio of the count rates on the fragment detector and the neutral detector and we notice that this ratio has a constant value of 0.88$\pm$0.02 for $t<10$\,ms. The deviation from unity can be ascribed to a small difference in the detection efficiencies of the two detectors.
For $t>10$\,ms the two detectors' count rates deviate strongly with the neutral detector rate being about five times that of the fragment detector at $t=10$\,s.
The right-hand vertical axis of the lower panel of Fig.~\ref{fig:Ag2_Ratios} is corrected for the slightly different detection efficiencies of the two detectors and thus directly gives the fraction of the decays that are due to fragmentation as a function of time.
It is clearly demonstrated that while fragmentation dominates initially, the contribution from electron detachment falls off more slowly with time and therefore becomes visible at around $t=10$\,ms and, in fact, dominates for $t>100$\,ms in the present experiment.

When interpreting decay curves like the ones in Fig.~\ref{fig:Ag2_Ratios}, it is important to emphasize that the large reduction in count rate by many orders of magnitude over the first few tens of milliseconds after injection does {\it not} mean that almost all stored ions have decayed. In fact a comparison of the numbers of injected ions and of detected fragments (taking into account geometrical limitations and detection efficiencies) indicate that about 99.9\,\% of the injected ions do {\it not} decay on the 10\,s time scale in the present experiment. Those ions, which are either stable or decay with very low decay rates due to lower internal energies can, however, still contribute to the neutral particle signal by residual-gas collisions. The three to four orders of magnitude reduction of the residual-gas density compared to room-temperature storage rings \cite{Schmidt2013} is therefore absolutely crucial for studies of the Ag$_2^-$ decays for seconds-long storage times. 

In pioneering work on silver dimer anion decays in the room-temperature storage ring ELISA \cite{Moller1997}, the neutral-particle signal could only be followed for 80\,ms due to the significant background from neutral products of collisions between the stored ion beam and the residual gas~\cite{Fedor2005}.
Our observation, that the count rate of neutrals is dominated by fragmentation early during storage, confirms the interpretation given in ref. \cite{Fedor2005}, namely that their signal is due to fragmentation of dimer 
anions. This also strongly supports their conclusion \cite{Fedor2005} that ions with very high angular momentum and with sufficient vibrational excitation for tunnelling through the centrifugal barrier are responsible for the fragmentation signal.

\begin{figure}
	\centering
	\includegraphics[width=\columnwidth]{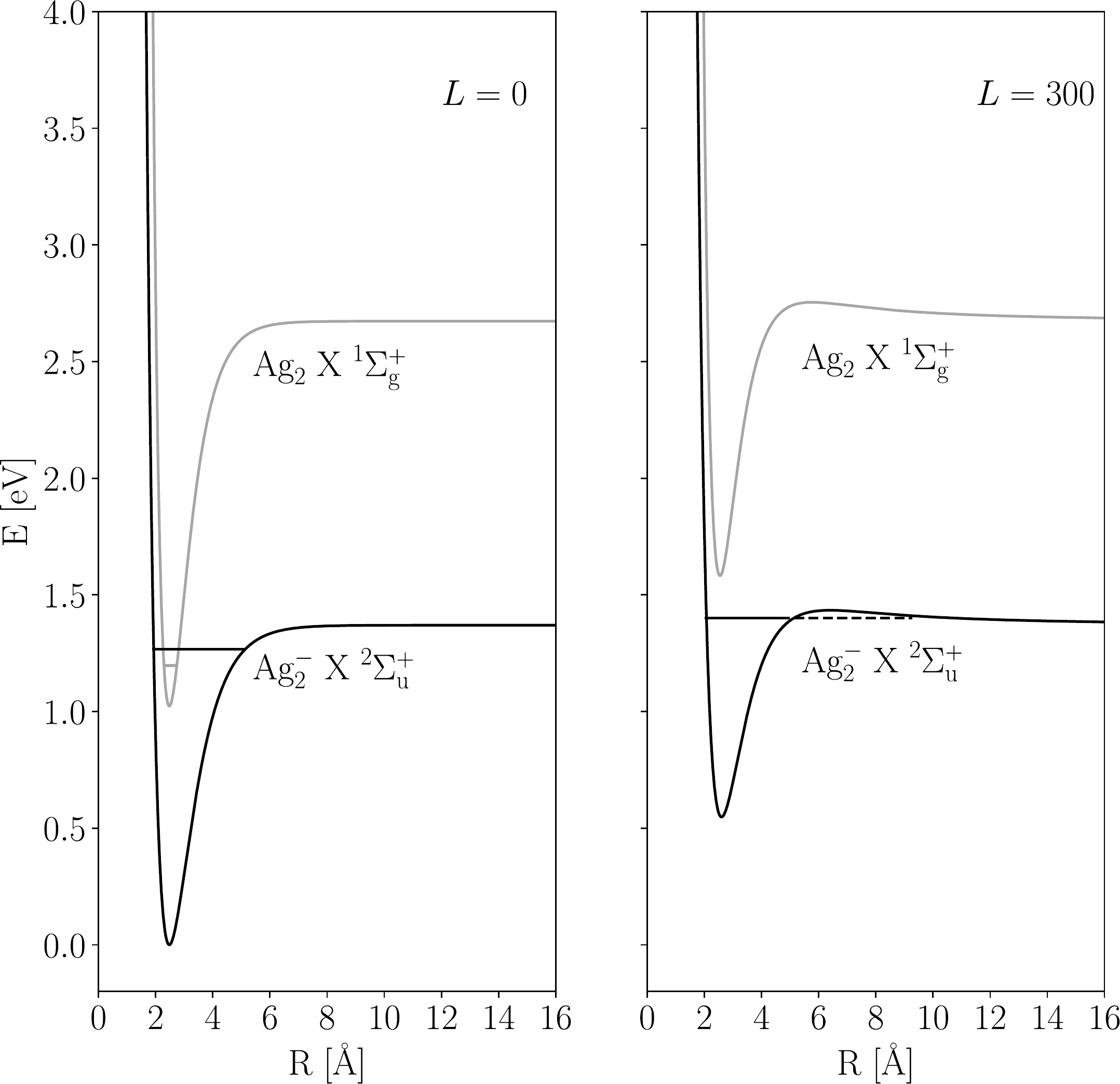}
	\caption{Approximate potential-energy curves based on literature values from \cite{Ho1990} and \cite{Morse1986} for rotational angular momentum $L=0$ (left panel) and $L=300$ (right panel). The black curves are the $L=0$ and $L=300$ dimer anion electronic ground state potentials, while the gray curves are the equivalent electronic ground state potentials of the neutral dimer. The observed fragmentation is interpreted as due to tunnelling through the centrifugal barrier for dimer anions in high angular-momentum levels as suggested in ref. \cite{Fedor2005}. The observed electron-detachment signal is ascribed to dimer anions in high vibrational levels with too low angular momenta to fragment but sufficient vibrational energy to detach the electron by the weak coupling of vibrational and electronic degrees of freedom (see text). The vibrational levels indicated in the left panel are $v=110$ for the anion and $v=7$ for the neutral dimer. The metastable rovibrational level illustrating the tunnelling fragmentation as an example in the right panel has $L=300$ and $v=88$.}
	\label{fig:curves}
\end{figure}

In Fig.~\ref{fig:curves}, we show approximate electronic ground state potential energy curves for the neutral silver dimer and for the dimer anion for non-rotating, $L=0$, systems (left panel) and for dimers with a high angular momentum $L=300$ (right panel). Rates for tunnelling through the centrifugal barrier, schematically indicated in the right panel of Fig~\ref{fig:curves} were calculated in \cite{Fedor2005}. These rates were found to depend very strongly on the vibrational and rotational quantum numbers of the large number of ro-vibrational levels, thereby leading to a quasi-continuum of decay rates and thus the power-law decays observed both here and in earlier experiments \cite{Fedor2005, Hansen2001}.

The potential energy curves shown in Fig.~\ref{fig:curves} are Morse potentials generated with parameter values from \cite{Ho1990} and \cite{Morse1986}.
Dimer anions occupying levels with modest angular momenta can accommodate high vibrational excitations and yet be energetically prohibited from fragmenting. As the electron affinity of  Ag$_2$ ({\it EA}$=1.023\pm 0.007$\,eV~\cite{Ho1990}) is smaller than the dissociation energy of  Ag$_2^-$ ($D_0=1.37 \pm 0.16$\,eV~\cite{Ho1990}), there are ro-vibrational levels for which electron emission is the only possible decay path, and the decay from such levels is therefore the only possible explanation for the observed electron emission. In the left panel of Fig.~\ref{fig:curves}, a single vibrational level ($v=110$) of the electronic ground state with $L=0$ is drawn as an example. For that particular level, energetically possible electron emission implies that the resulting neutral dimer has a vibrational quantum number of $v\,\leq\,10$. This illustrates that the energy transferred to the electron is initially stored as vibrational excitation in the dimer anion and that the emission of the electron is accompanied by a sudden, dramatic, decrease of the vibrational excitation. The total center-of-mass kinetic energy of the two nuclei instantaneously decreases by an amount of at least 1.023$\pm$0.007\,eV~\cite{Ho1990} corresponding to the dimer electron affinity.
As no diabatic potential-energy curve crossings are available to drive the energy transfer from the nuclear to the electronic degrees of freedom, this is clearly a most dramatic departure from a Born-Oppenheimer description of the dynamics.

\begin{figure}
	\centering
	\includegraphics[width=0.9\columnwidth]{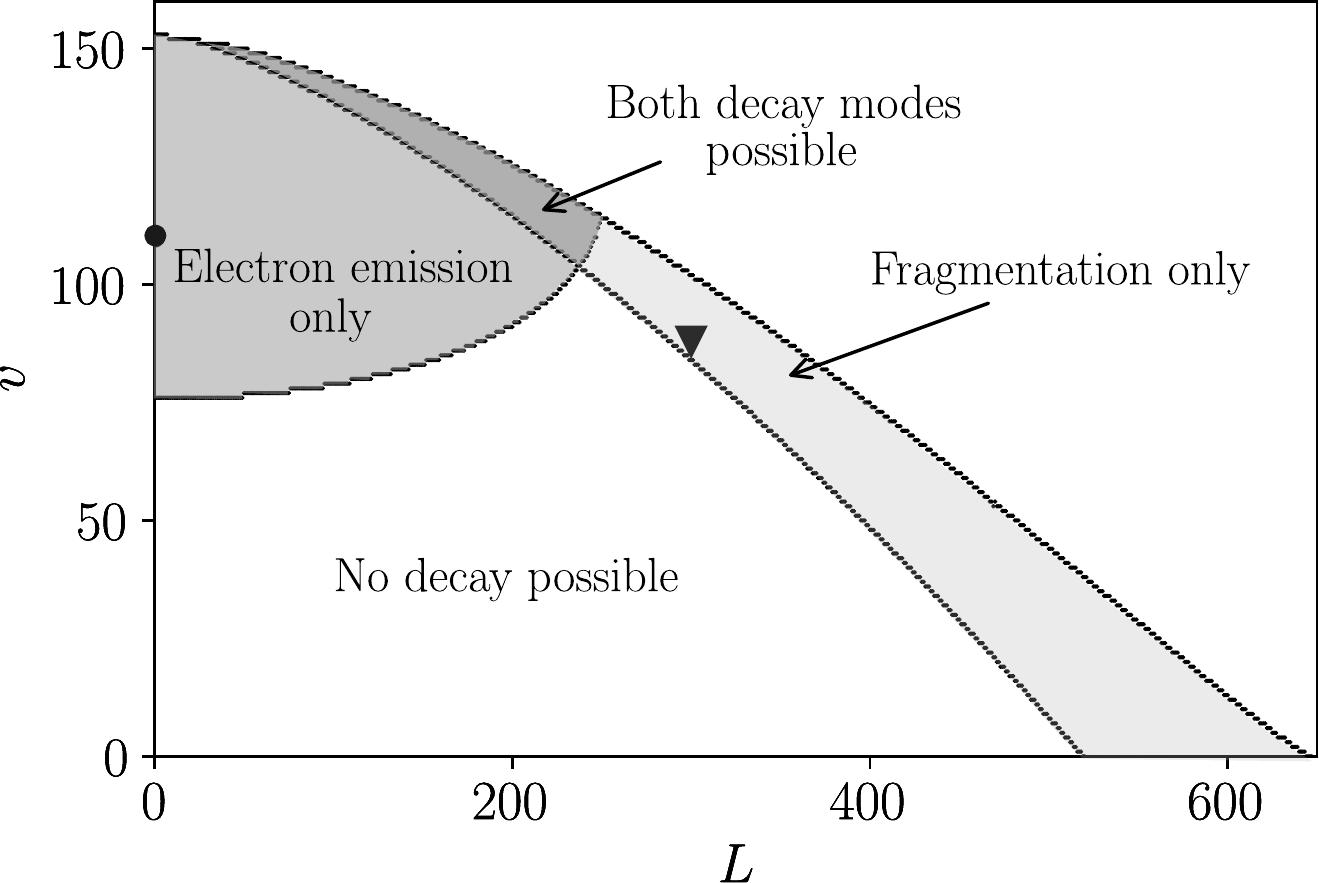}
	\caption{A map of the ro-vibrational levels of Ag$_2^-$ indicating for which combinations of rotational and vibrational excitations fragmentation and/or electron emission are energetically allowed. The dot and triangle markings indicate the $L$ and $v$ corresponding to the levels drawn in the left and right panels of Fig. \ref{fig:curves}, respectively.} 
	\label{fig:Lv_map}
\end{figure}
As mentioned above, the ions decaying by fragmentation have, in part, different ro-vibrational excitations than those decaying by electron emission. In order to investigate this in more detail, we calculate the ro-vibrational energy level structure of neutral and anion silver dimers. We again use the Morse potentials based on parameters from refs. \cite{Ho1990} and \cite{Morse1986}, and for each value of angular momentum, $L$, we solve the radial Schr\"odinger equation numerically to find the vibrational levels as the eigenstates for that given potential including the centrifugal distortion. The energy of each ro-vibrational level is then compared to both the energy of the separated Ag$^-$ ion and Ag atom and the lowest vibrational level of the neutral dimer with the same value of $L$ as the dimer anion. The emitted electron can only carry away a few quanta of angular momentum and here the angular momentum of the electron is neglected. This way we determine---for each rovibrational level---whether it is stable against fragmentation and whether it can decay by electron emission. 
In Fig. \ref{fig:Lv_map}, we show a map over the ro-vibrational levels of Ag$_2^-$ indicating whether they are stable or if they can decay by fragmentation, electron emission or by both these processes.
The dimer anions are formed in the ion source in violent collisions of 6\,keV Cs$^+$ ions on a silver surface and have been found to occupy a wide range of ro-vibrational levels \cite{Wucher1996}.
Indeed, the presently observed decays and the large component of stable (or extremely long-lived) Ag$_2^-$ anions suggest a very broad ro-vibrational distribution.

We argued above that the dramatic decrease in vibrational excitation accompanying the electron emission is particularly remarkable as no avoided crossing of the two involved potential energy curves is available to drive the transition.
In a quantum mechanical description of electron emission, the non-adiabatic coupling between the anion and the ionization continuum will give rise to coupling elements of the form 
\textendash$\hbar^2$/(2$\mu$)$\langle\chi_{v'L'}^i$\textbar$f_{ij}(R)$d/d$R$\textbar$\chi_{vL}^j\rangle$,
where $\mu$ is the reduced mass of the nuclei and $\chi^i_{v'L'}$ and $\chi^j_{vL}$ are the nuclear wave functions of a given ro-vibrational state of Ag$_2$ and Ag$_2^-$, respectively.
The non-adiabatic coupling element, $f_{ij}(R)$, is strictly zero in the Born-Oppenheimer approximation, but here we formally express it as
$f_{ij}(R)$=$\langle\psi_i$\textbar{$\partial$/$\partial$}$R$\textbar$\psi_j\rangle$, where $\psi_i$ and $\psi_j$ are the full electronic wave functions of the neutral plus a free electron and of the anion, respectively.
The non-adiabatic coupling is in general large when the adiabatic potential energy curves are close in energy. Here the potentials of the Ag$_2^-$ and Ag$_2$ are far apart and as a result $f_{ij}$ will be small (but not zero) and not a strong function of $R$.

In spite of this weak coupling, the electron emission process dominates the measured decay signal for $t>$100\,ms. In fact, 
electron emission events account for 13\,\% of the total number of decay events during the 10\,s observation time.
This surprising observation relies on the exceptional experimental conditions and the absence of any other decay channels rather than on the efficiency of the electron emission process. To quantify the latter, we use the detailed balance approach \cite{Weisskopf1937, Hansen2013} to estimate the cross section for the time-reversed process---the attachment of a free electron to the neutral dimer via a strong increase of the vibrational excitation of the dimer system. Assuming an electron energy of 20\,meV (a typical vibrational spacing) and a decay rate of 10\,s$^{-1}$, which is characteristic of the measured time scale, indicate an extremely small electron attachment cross section on the order of 10$^{-9}$~\AA$^2$.
This semi-empirical cross section estimate is about nine orders of magnitude smaller than the geometrical cross section of the dimer, which illustrates the weakness of the coupling. We note that the analysis leading to figures \ref{fig:curves} and \ref{fig:Lv_map} implies that electron attachment is impossible to neutral dimers in high rotationally excited levels ($L'\textgreater 250$).

In conclusion, we have investigated the decay of internally hot silver dimer anions stored in a cryogenic electrostatic ion-storage ring with exceptionally low residual-gas density. By separately detecting neutral and charged products of the decay, we demonstrate experimentally that fragmentation is the dominating decay process for the initial parts of the decay (up to 100\,ms of storage). Further, we unambiguously identify a prominent contribution from electron emission. This is remarkable as no potential energy curve crossings are available to mediate a coupling between the nuclear and electronic degrees of freedom. In the absence of of such potential-energy curve crossings, this process represents a dramatic example of a breakdown of the Born-Oppenheimer approximation.

\section*{Acknowledgements}
This work was performed at the Swedish National Infrastructure, DESIREE (Swedish Research Council contract No. 2017-00621). Furthermore, HC, HZ and HS thank the Swedish Research Council for individual project grants (Nos. 2015-04990,  2016-04181, 2018-04092). A.S. acknowledges support by the Austrian Science Fund FWF within the DK-ALM (W1259-N27).


\end{document}